%
%
\def\Cov{{\rm Cov}}

\def\Var{{\rm Var}}

\catcode`\@=11
\def\xtitleb#1#2{\par\stepc{Tm}
    \resetcount{Tn}
    \if N\lasttitle\else\vskip\tbbeforeback\fi
    \bgroup
       \normalsize
       \raggedright
       \pretolerance=10000
       \it
       \setbox0=\vbox{\vskip\tbbefore
          \normalsize
          \raggedright
          \pretolerance=10000
          \noindent \it #1.\arabic{Tm}.\ \ignorespaces#2
          \vskip\tbafter}
       \dimen0=\ht0\advance\dimen0 by\dp0\advance\dimen0 by 2\baselineskip
       \advance\dimen0 by\pagetotal
       \ifdim\dimen0>\pagegoal
          \ifdim\pagetotal>\pagegoal
          \else \if N\lasttitle\eject\fi \fi\fi
       \vskip\tbbefore
       \if N\lasttitle \penalty\subsection@penalty \fi
       \global\subsection@penalty=-100
       \global\subsubsection@penalty=10007
       \noindent #1.\arabic{Tm}.\ \ignorespaces#2\par
    \egroup
    \nobreak
    \vskip\tbafter
    \let\lasttitle=B%
    \parindent=0pt
    \everypar={\parindent=\stdparindent
       \penalty\z@\let\lasttitle=N\everypar={}}%
       \ignorespaces}
\def\xautnum#1{\global\advance\eqnum by 1\relax {\rm (#1\the\eqnum)}}
\catcode`\@=13
%
%
%
  \MAINTITLE={ Improving the accuracy of mass reconstructions from
  weak lensing: from the shear map to the mass distribution }
  \SUBTITLE={ ????? }  \AUTHOR={ Marco Lombardi and Giuseppe Bertin }
  \OFFPRINTS={ M. Lombardi } \INSTITUTE={ Scuola Normale Superiore,
  Piazza dei Cavalieri 7, I 56126 Pisa, Italy } \DATE={ ????? }
  \ABSTRACT={ In this paper we provide a statistical analysis of the
  parameter-free method often used in weak lensing mass
  reconstructions. It is found that a proper assessment of the errors
  involved in such a non-local analysis requires the study of the
  relevant two-point correlation functions. After calculating the
  two-point correlation function for the reduced shear, we determine
  the expected error on the inferred mass distribution and on other
  related quantities, such as the total mass, and derive the error
  power spectrum. This allows us to optimize the reconstruction
  method, with respect to the kernel used in the inversion
  procedure. The optimization process can also be carried out on the
  basis of a variational principle. In particular, we find that
  curl-free kernels are bound to lead to more accurate mass
  reconstructions. Our analytical results clarify the arguments and
  the numerical simulations by Seitz \& Schneider (1996). }
  \KEYWORDS={ gravitational lensing -- dark matter -- galaxies:
  clustering } \THESAURUS={ 12.07.1; 12.04.1; 11.03.1 } \maketitle
  \MAINTITLERUNNINGHEAD{ Accurate mass from weak lensing }
  \AUTHORRUNNINGHEAD{ M. Lombardi \& G. Bertin }
%
%
\titlea{Introduction}
One of the most interesting applications of gravitational lenses is
the determination of the projected mass distribution from weak lensing
observations. As noted, among others, by Webster (1985), the mean
orientation of a large number of distant galaxies gives a measure of
the {\it shear\/} associated with the lens. The observed shear can
then be used to derive the two-dimensional mass distribution of the
lens responsible for the deformation induced on the background. This
last step can be carried out in two different ways. The easier route
is to use a specific model for the lens with a number of free
parameters that will be determined by a comparison between the
observed and the predicted shear (see, e.g., Kneib {\it et al}.
1996). A more general procedure is the so called ``parameter-free
reconstruction'' (Kaiser \& Squires 1993; see also Bartelmann {\it et
al}.\ 1996). In this latter method the mass distribution can be
directly determined from the shear map, provided that the shear is
known with sufficient accuracy and detail, which requires the
existence of a large number of source galaxies.

Such reconstruction techniques are, of course, a powerful tool to
study the matter distribution in clusters (see e.g.\ Tyson, Valdes,
Wenk 1990, Fahlman {\it et al.} 1994, Smail {\it et al.} 1994) and for
large scale structures. It is then important to optimize the
reconstruction process in order to make the best use of the
observations. For this purpose, we have to assess the expected error
of a specific reconstruction method, which is the main goal of the
present paper.

In this article we focus our attention on the parameter-free method,
mainly because this is more general and does not depend on the
particular lens under consideration. In a previous paper (Lombardi \&
Bertin 1998, hereafter Paper~I) we have provided expressions for the error
involved in the {\it local\/} measurements of the shear (or the
reduced shear) of the lens as a function of the parameters
characterizing the distribution of source galaxies. Here we extend the
statistical analysis to the inferred global mass distribution.

The text is organized as follows. In Sect.~2 we introduce the spatial
weight function and we briefly describe various reconstruction methods
used to infer the lens mass distribution. In Sect.~3 we calculate the
expected error on the measured shear in the regime of weak lensing and
in the more general case as a function of position in a given field of
the sky; here the formulae of Paper~I are generalized to the two-point
correlation function for the shear map (see Eq.~(25)). This important
result is then used in Sect.~4 to calculate the expected errors on the
mass distribution associated with the various reconstruction
methods. The results are then compared, in Sect.~5, to the simulations
by Seitz \& Schneider (1996).

The main result of the paper is contained in Eq.~(35) (together with
Eq.~(26)) that describes the two-point correlation function for the
mass density $\kappa$ obtained from weak lensing analysis. This proves
that, in order to optimize the reconstruction process for observations
in a finite area of the sky, a curl-free kernel should be used (see
Eq.~(36)). This behavior is confirmed by numerical simulations.

\titlea{From the shear map to the mass distribution}
We consider a field of the sky with $N$ source galaxies located at
$\vec{\theta}^{(n)}$ and characterized by observed quadrupole
$Q^{(n)}$ and ellipticity $\chi^{(n)}$ (see Appendix~A for a summary
of the adopted notation). Here we suppose that the galaxies are
observed inside a field $\Omega$ of area $A$, with mean spatial
density equal to $\rho = N / A$. In the rest of the paper we will
reserve the term ``weak lensing'' to the limit of small lens density
$\kappa \ll 1$, i.e.\ $\gamma \simeq g$.

\titleb{Spatial weight function}
Source galaxies located close to a given position $\vec{\theta}$ will
better constrain the value of the reduced shear $g(\vec{\theta})$ at
such location. In order to describe this effect, we may thus introduce
a suitable weight function $W(\vec{\theta}, \vec{\theta}')$. The first
argument of the weight function, $\vec{\theta}$, represents the point
of the sky under consideration and for which we want to measure the
shear $g(\vec{\theta})$, while the second argument $\vec{\theta}'$
represents the location of one observed galaxy. The weight function
should penalize galaxies far from $\vec{\theta}$, i.e.\
$W(\vec{\theta}, \vec{\theta} + \vec{\vartheta})$ should decrease for
increasing $\| \vec{\vartheta} \|$. Some additional ``natural''
conditions can be given to further characterize a specific choice of
weight function and these are most convenient when applied (beginning
with Eq.~(22)) to a spatially continuous distribution of source
galaxies with density $\rho$. Here we list a few possible assumptions,
where the first is obviously the least restrictive:
\medskip
\item{1.} The weight function is {\it even} with respect to
$\vec{\vartheta}$, i.e.
$$
W(\vec{\theta}, \vec{\theta} + \vec{\vartheta}) = W(\vec{\theta},
\vec{\theta} - \vec{\vartheta}) \; .
\eqno \autnum
$$
\item{2.} The weight function is said {\it invariant upon translations\/},
if it is {\it even\/} (see above) and if it depends only on the
difference $\vec{\theta} - \vec{\theta}'$:
$$
W(\vec{\theta}, \vec{\theta}') = W(\vec{\theta} - \vec{\theta}') \; .
\eqno \autnum
$$
\item{3.} One natural choice is that of a {\it Gaussian\/} dependent
only on the distance $\| \vec{\theta} - \vec{\theta}' \|$,
$$
W(\vec{\theta}, \vec{\theta}') = {1 \over 2 \pi \rho \sigma_W^2} \exp
\left(-  {\| \vec{\theta} - \vec{\theta}' \|^2  \over 2 \sigma_W^2}
\right) \; , 
\eqno \autnum
$$
where the angular scale $\sigma_W$ should be sufficiently large to
ensure the presence of an adequate number of galaxies in a disk of
radius $\sigma_W$ centered on the generic point $\vec{\theta}$.
\medskip
The value of the weight function at a given point $\vec{\theta}'$, of
course, has no particular meaning: only {\it relative\/} values are
significant. Indeed all the following results can be shown to be
unaffected if we merely multiply the weight function by a
constant. Thus, we may always choose a {\it normalized\/} weight
function, so that
$$
\sum_{n=1}^N W \bigl( \vec{\theta}, \vec{\theta}^{(n)} \bigr) = 1
\eqno \autnum
$$
for every $\vec{\theta}$. Of course, such a normalization will remove
the translation invariance property, if initially present in
$W$. Still, the invariance may be retained when we will move to a
continuous description (see comment after Eq.~(22)).

The spatial weight function operates much like the ``shape''
weight functions considered in Paper~I (see Eqs.~(23) and (21)
there). In particular, using the isotropy condition, we can obtain the
shear map $g(\vec{\theta})$ either from
$$ 
\sum_{n=1}^N W\bigl( \vec{\theta}, \vec{\theta}^{(n)} \bigr) \chi^{\rm s} 
\bigl( \chi^{(n)}, g(\vec{\theta}) \bigr) = 0 \; , 
\eqno \autnum
$$
or from
$$
\chi^{\rm s} \biggr( \sum_{n=1}^N W \bigl( \vec{\theta}, 
\vec{\theta}^{(n)} \bigr) Q^{(n)}, g(\vec{\theta}) \biggr) = 0 \; . 
\eqno \autnum
$$
In Paper~I we discussed the different merits of the two options. In
the limit of ``sharp'' distributions for the source galaxies ($c \ll
1$), they both lead to the same determination of the true reduced
shear $g_0$ (apart from the ambiguity associated with the $g \mapsto
1/g^*$ invariance).

As we will see the angular scale of the weight function $W$, i.e.\ the
diameter of the set where $W(\vec{\theta}, \vec{\theta}')$ is
significantly different from zero, determines a lower bound for the
smallest details shown in the reconstructed map
$\kappa(\vec{\theta})$. For example, in the weak lensing limit and for
a weight function invariant upon translations, the mean value of the
measured density $\kappa$ is related to the true density $\kappa_0$
through the expression (see Appendix~C)
$$
\langle \kappa (\vec{\theta}) \rangle = \rho \int W(\vec{\theta} -
\vec{\theta}') \kappa_0(\vec{\theta}') \, \diff^2 \theta'\; .
\eqno \autnum
$$
This relation, similar to that found when an image is degraded by a
PSF, suggests the possible application of deconvolution techniques
(see Lucy 1994) to the present context.

\titleb{Weak lensing regime}
We first consider the case where $\Omega$ is very large, so that the
field is identified with the whole plane (the effects of the
boundaries will be discussed soon, in Sect.~2.4).

There are basically two ways to reconstruct the mass distribution
$\kappa(\vec{\theta})$ from the shear map $g(\vec{\theta})$ (Seitz \&
Schneider 1996). The first, more natural method is based on the
integral relation
$$
\kappa(\vec{\theta}) = \int D_i(\vec{\theta} - \vec{\theta}') \gamma_i( 
\vec{\theta}' ) \, \diff^2 \theta' \; ,
\eqno \autnum
$$
where the kernel $D_i$ is given by (Kaiser \& Squires 1993)
$$
\left( \matrix{D_1(\vec{\theta}) \cr D_2(\vec{\theta})} \right) = {1
\over \pi \| \vec{\theta} \|^4} \left(  \matrix{ \theta_1^2 -
\theta_2^2 \cr 2 \theta_1 \theta_2 \cr} \right) \; .
\eqno \autnum
$$
In the weak lensing limit $\gamma \simeq g$, and thus the reduced
shear map can be used directly in Eq.~(8) to derive
$\kappa(\vec{\theta})$. Note that the inverse relation holds with the
{\it same\/} kernel
$$
\gamma_i(\vec{\theta}) = \int D_i(\vec{\theta} - \vec{\theta}') \kappa
( \vec{\theta}' ) \, \diff^2 \theta' \; .
\eqno \autnum
$$
This relation will turn out to be useful in Appendix~C.

A second possibility, which can be proved to be mathematically
equivalent to the first, is based on the exact relation (see Kaiser
1995)
$$
\nabla \kappa = \vec{u}(\theta) = - \left( \matrix{ \gamma_{1,1} +
\gamma_{2,2} \cr 
\gamma_{2,1} - \gamma_{1,2}} \right) \; ,
\eqno \autnum
$$
which is a direct consequence of the thin lens equations. Here
$\gamma_{i,j} = \partial \gamma_i / \partial \theta_j$. By analogy with
the condition used to derive Eq.~(8), if we assume that
$\kappa(\vec{\theta})$ vanishes for large values of $\| \vec{\theta}
\|$, Eq.~(11) can be inverted to give
$$
\kappa(\vec{\theta}) = \int H^{\rm KS}_i(\vec{\theta} - \vec{\theta}')
u_i(\vec{\theta}') \, \diff^2 \theta' \; ,
\eqno \autnum
$$
with the kernel
$$
H^{\rm KS}_i(\vec{\theta}) = {\theta_i \over 2 \pi \| \vec{\theta}
\|^2} \; .
\eqno \autnum
$$
In Eq.~(12) the shear map enters through the vector $\vec{u}$, which,
in the weak lensing limit, involves the derivatives of
$g(\vec{\theta})$. This second method thus introduces undesired
differentiations, but it has the advantage that it is more easily
generalized to include the effects of the boundaries (see Sect.~2.4
below). 

\titleb{The general case}
When the lens is not weak, Eq.~(8)
$$
\kappa(\vec{\theta}) = \int D_i(\vec{\theta} - \vec{\theta}') \bigl( 1
- \kappa(\vec{\theta}') \bigr) g_i( \vec{\theta}' ) \, \diff^2
\theta'
\eqno \autnum
$$
can be solved by iteration (Seitz \& Schneider 1995).

The second method, related to Eq.~(12), has been generalized by Kaiser
(1995) for strong lenses. If we introduce
$\tilde{\kappa}(\vec{\theta}) = \ln \bigl( 1 - \kappa(\vec{\theta})
\bigr)$ and the new vector
$$
\tilde{u}_i = {1 \over 1 - |g|^2} \left( \matrix{ 1 + g_1 & g_2 \cr
g_2 & 1 - g_1 } \right) \left( \matrix{ g_{1,1} + g_{2,2} \cr
g_{2,1} - g_{1,2} } \right) \; ,
\eqno \autnum
$$
then it is possible to show that the relation $\nabla
\tilde{\kappa}(\vec{\theta}) = \tilde{\vec{u}}(\vec{\theta})$
holds. As a result $\tilde{\kappa}$ can be obtained from
$\tilde{\vec{u}}$ via the same integral equation (12) used earlier. The
fact that $\tilde{\kappa}$ is determined only up to a constant here
translates into a non-trivial invariance for the density distribution
$\kappa(\vec{\theta})$, under the transformation (see Schneider \&
Seitz 1995)
$$
\kappa(\vec{\theta}) \longmapsto (1 - C) \kappa(\vec{\theta}) + C \; ,
\eqno \autnum
$$ 
consistent with Eq.~(14).

\titleb{Effect of the boundaries}
The methods described so far assume an infinite domain of
integration. In practice, one can measure the shear only in a finite
area (e.g.\ the CCD area), which is often small compared to the angular
size of the lensing cluster. Therefore, the relations given earlier
should be properly modified.

We briefly noted that the second method is better suited for the
purpose. In the following, for simplicity, we consider only the weak
lensing limit, but the equations that we will provide can be easily
extended to the general case by replacing $(\vec{u},
\kappa) \mapsto (\tilde{\vec{u}}, \tilde{\kappa})$. The relations
suggested by Seitz \& Schneider (1996) for mass reconstruction in a
field $\Omega$ of finite area $A$ are of the form
$$
\kappa(\vec{\theta}) - \bar{\kappa} = \int_\Omega G_i(\vec{\theta},
\vec{\theta}') u_i(\vec{\theta}') \, \diff^2
\theta' 
\; .
\eqno \autnum
$$
Here $\bar{\kappa}$ is a constant representing the average of
$\kappa$, while $\vec{G}$ is a suitable kernel. The kernel is
chosen so as to give the correct mass distribution if $\vec{u}$ could
be measured with no errors (see Eq.~(C14)). There is however some
freedom left in the choice of the kernel, mainly because it returns a
scalar field ($\kappa$) from a vector field ($\vec{u}$). This freedom
will be further discussed later on. One interesting
kernel, called {\it noise filtering}, has been introduced by Seitz \&
Schneider (1996)
$$
\vec{H}^{\rm SS}(\vec{\theta}, \vec{\theta}') =
-\nabla_{\vec{\theta}'} {\cal H}^{\rm SS}(\vec{\theta}, \vec{\theta}')
\; , 
\eqno \autnum
$$
where ${\cal H}^{\rm SS}$ is the solution of Neumann's boundary
problem ($\vec{n}$ is the unit vector orthogonal to $\partial \Omega$)
$$\eqalignno{
& \nabla^2_{\vec{\theta}'} {\cal H}^{\rm SS}(\vec{\theta},
\vec{\theta}') =  \delta(\vec{\theta} - \vec{\theta}') - {1 \over A}
\; , & \autnum \cr
& {\partial {\cal H}^{\rm SS}(\vec{\theta},\vec{\theta}') \over
\partial \theta'_i} n_i(\vec{\theta'}) = 0 \qquad \forall
\vec{\theta}' \in \partial \Omega \; . & \autnum \cr}
$$
The term related to the area $A$ ensures the proper applicability of
the Gauss theorem. Note that the kernel $\vec{H}^{\rm SS}$ has
vanishing curl, i.e.
$$
\nabla_{\vec{\theta}'} \wedge \vec{H}^{\rm SS}(\vec{\theta},
\vec{\theta}') = 0 \; .
\eqno \autnum
$$

\titlea{Measurements of the reduced shear map and of the two-point
correlation function} 

In this section we will give an expression for the reduced shear map
measured using Eqs.~(5) or (6). In Paper~I we have calculated the
statistics associated with a {\it local\/} shear measurement under the
hypothesis that the probability distribution for the source
ellipticity $\chi^{\rm s}$ is {\it sharp}, i.e.\ most source galaxies
are nearly round. Now we consider situations where the reduced shear
is a function of the position $\vec{\theta}$, but we assume that
$g(\vec{\theta})$, a smooth function of $\vec{\theta}$, does not
change significantly on the angular scale $\sigma_W$ of the weight
function $W(\vec{\theta},
\vec{\theta}')$. An important new aspect of the analysis that has to
be addressed here, in view of the goal of determining the error on the
reconstructed mass, is the calculation of the two-point correlation
function for the shear map.

So far we have considered source galaxies with random orientation but
with fixed position on the sky (corresponding to $\vec{\theta}^{(n)}$
on the observer's sky). It is interesting to average all results by
assuming that galaxies have random positions. The result of the
average can be approximated by considering a continuous distribution
of galaxies with density $\rho$ (number of galaxies per
steradian). This leads us to ignore, for the moment, the effects of
Poisson noise associated with the finite number of source galaxies
(further comments are given at the end of Sect.~4.1). Following Seitz
\& Schneider (1996), we consider a homogeneous distribution of galaxies in
$\vec{\theta}$, i.e.\ in the {\it observer's plane}. If $\rho$ is
independent of $\vec{\theta}$, we may change summations with integrals
using the rule $\sum_n \mapsto \rho
\int \diff^2 \theta'$. Here, to simplify the derivations and the
discussion, we adopt, for every $\vec{\theta}$, the normalization
$$
\rho \int W(\vec{\theta}, \vec{\theta}') \, \diff^2 \theta' = 1
\eqno \autnum
$$
for the weight function. For an infinite field $\Omega$ or for the
case described in Fig.~1, this normalization does not break the
translation invariance of $W$. Then as shown in Appendix~B, the
relation between expected and true value of $g$, corresponding to
Eq.~(7), is
$$
\bigl\langle g(\vec{\theta}) \bigr\rangle = \rho \int W(\vec{\theta},
\vec{\theta}') g_0(\vec{\theta}') \, \diff^2 \theta' \; .
\eqno \autnum
$$
As is intuitive, ``near'' galaxies give the most important
contribution to the measured value of $g$.

The correct generalization of the covariance matrix $\Cov_{ij}(g)$
when $g$ is a function of the position $\vec{\theta}$ is a two-point
correlation function:
$$\eqalignno{
& \Cov_{ij}(g; \vec{\theta}, \vec{\theta}') = & \cr 
& \quad \bigl\langle \bigl( g_i(\vec{\theta}) -
\overline{g_i(\vec{\theta})} \bigr) \bigl( g_j(\vec{\theta}') -
\overline{g_j(\vec{\theta}')} \bigr) \bigr\rangle . &
\autnum \cr}
$$
Note that the knowledge of the ``diagonal'' values $\Cov_{ij}(g;
\vec{\theta}, \vec{\theta})$ is not sufficient to calculate the error
on other variables, such as the density distribution $\kappa$,
determined from $g$ (cf.\ Eq.~(32) and Eq.~(35)).

If we assume that the weight function is {\it even\/} (property 1 of
Sect.~2.1), then the two-point correlation function of $g$ can be
written in the simple form (see Appendix~B)
$$\eqalignno{
&\Cov_{ij}(g; \vec{\theta}, \vec{\theta}') = {c \over 4} \bigl( 1 -
\bigl| \langle g(\vec{\theta}) \rangle \bigr|^2 \bigr) \bigl( 1 -
\bigl| \langle g(\vec{\theta}') \rangle \bigr|^2 \bigr) \delta_{ij}
\times {} & \cr  
& \quad {} \times \rho \int W(\vec{\theta}, \vec{\theta}'')
W(\vec{\theta}', \vec{\theta}'') \, \diff^2 \theta'' \; .
& \autnum \cr}
$$
Here $c$ is the covariance of the ellipticity distribution of the
source galaxies. In this equation, as noted for Eq.~(23), we suppose
the weight function $W(\vec{\theta}, \vec{\theta}')$ to be
normalized.

In the weak lensing limit Eq.~(25) then reduces to
$$\eqalignno{
\Cov_{ij}(\gamma; \vec{\theta}, \vec{\theta}') & = {c \rho \delta_{ij}
\over 4} \int W(\vec{\theta}, \vec{\theta}'') W(\vec{\theta}',
\vec{\theta}'') \, \diff^2 \theta'' & \cr
& = {c \delta_{ij} \over 16 \pi \rho \sigma_W^2} \exp \left( - {\|
\vec{\theta} - \vec{\theta}' \|^2 \over 4 \sigma^2_W} \right) \; . &
\autnum \cr}
$$
The last relation holds for a Gaussian weight function of the form
given in Eq.~(3); here $\Cov_{ij}(\gamma; \vec{\theta},
\vec{\theta}')$ is a simple Gaussian with variance $2 \sigma^2_W$ and
depends only on $\| \vec{\theta} - \vec{\theta}' \|$. The variance
$\bigl\langle \bigl( \gamma_i(\vec{\theta}) - \overline{\gamma_i(\vec{\theta})}
\bigr)^2 \bigr\rangle$ of $\gamma_i(\vec{\theta})$ is simply
$\Cov_{ii}(\gamma; \vec{\theta}, \vec{\theta}) = c / (16 \pi \rho
\sigma_W^2)$ (without summation on $i$) and thus increases if
$\sigma_W$ decreases. This behavior can be explained by considering that the
number of galaxies used for a single point is of the order of $\rho
\sigma_W^2$. Notice also that $\sigma_W$ sets the scale length of the
covariance of $\gamma$: measurements of $\gamma(\vec{\theta})$ and
$\gamma(\vec{\theta}')$ are uncorrelated if $\| \vec{\theta} -
\vec{\theta}' \| \gg \sigma_W$.

\titlea{Measurements of the mass distribution}
It is not difficult, at least in principle, to calculate the error on
$\kappa(\vec{\theta})$ from the two-point correlation function of $g$.
The error on $\kappa$, of course, depends on the reconstruction method
used. For this reason, following Sect.~2, we consider different
methods separately. For simplicity, we suppose that the weight
function $W$ is invariant upon translations. Moreover, we suppose that
the angular scale of the weight function $W$ is much smaller than the
angular scale of $\kappa$ (i.e.\ the scale where $\kappa$ varies
significantly). In general, if we ignore edge effects, the relation
between the error on $g$ and that on $\kappa$ is given by
$\Cov_{ij}(\gamma; \vec{\theta},
\vec{\theta}') = \delta_{ij} \Cov(\kappa; \vec{\theta},
\vec{\theta}')$. Here we show only the results obtained, referring to
Appendix~C for a derivation. As in Sect.~2, we first refer to the case
where $\Omega$ is identified with the whole plane (finite field
effects will be addressed in Sect.~4.3 below).

\titleb{Weak lensing regime}
In this case we can use either Eq.~(8) or Eq.~(12). A rather
surprising result is that both methods lead to the same mean values
and errors for $\kappa$. The result for the mean value has already
been stated in Eq.~(7), i.e.\ the measured mass distribution
$\bigl\langle \kappa(\vec{\theta}) \bigr\rangle$ is the convolution
of the weight function $W$ with the true mass distribution
$\kappa_0(\vec{\theta})$. 

For an ``isolated lens'' (a case where $\kappa$ is taken to vanish
outside a certain domain $\Omega_{\rm in}$) the ambiguity associated
with Eq.~(16) is resolved and the concept of total mass of the lens
becomes meaningful. In the weak lensing limit, from any reconstructed
$\kappa(\vec\theta)$ one can in principle accept also
$\kappa_C(\vec\theta) = \kappa(\vec\theta) + C$. Now if we know that
the density vanishes outside $\Omega_{\rm in}$, the constant $C$ can
be determined by requiring
$$
\int_{\Omega_{\rm out}} \!\! \diff^2\theta \, \kappa_C(\vec\theta) = 0
\; , 
\eqno \autnum
$$
where $\Omega_{\rm out} = \Omega \setminus \Omega_{\rm in}$ is the
part of the field not contained in $\Omega_{\rm in}$. Therefore, the
appropriate density to be used is
$$
\kappa_C(\vec\theta) = \kappa(\vec\theta) - {1 \over A_{\rm out}}
\int_{\Omega_{\rm out}} \!\! \kappa(\vec\theta') \, \diff^2 \theta' \;
. 
\eqno \autnum
$$
The associated total mass is
$$\eqalignno{
M & = \int_\Omega \kappa_C(\vec\theta) \, \diff^2\theta =
\int_{\Omega_{\rm in}} \! \kappa_C(\vec\theta) \, \diff^2 \theta & \cr
& = \int_{\Omega_{\rm in}} \! \kappa(\vec\theta) \, \diff^2\theta -
{A_{\rm in} \over A_{\rm out}} \int_{\Omega_{\rm out}} \!\!
\kappa(\vec\theta) \, \diff^2\theta \; . & \autnum \cr}
$$
Therefore:
$$\eqalignno{
\langle M \rangle & = \int_{\Omega_{\rm in}} \! \bigl\langle
\kappa(\vec{\theta}) \bigr\rangle \, \diff^2 \theta - {A_{\rm in}
\over A_{\rm out}} \int_{\Omega_{\rm out}} \!\! \bigl\langle
\kappa(\vec\theta) \bigr\rangle \, \diff^2 \theta & \cr 
& = \int_{\Omega_{\rm in}} \! \kappa_0(\vec{\theta}) \, \diff^2 \theta
= M_0 \; , & \autnum \cr} 
$$
where $M_0$ is the true mass of the lens. In other words, the {\it
smoothing effect\/} associated with $W$ does not change the measured
total mass $M$ of the lens.

The covariance of the lens distribution $\kappa$ can be shown to be
equal to (for both Eqs.~(5) and (6)):
$$
\Cov (\kappa; \vec{\theta}, \vec{\theta}') = {c \rho \over 4} \int
W(\vec{\theta}, \vec{\theta}'') W(\vec{\theta}', \vec{\theta}'') \,
\diff^2 \theta'' \; . 
\eqno \autnum
$$
In comparing Eq.~(31) to Eq.~(26), one should note that the similarity
of results refers statistically to average errors, but not to the
individual errors of one reconstruction. The variance in the measure
of the total mass is the integral of the covariance of $\kappa$:
$$\eqalignno{
\Var(M) & = \int_{\Omega_{\rm in}} \! \diff^2 \theta \int_{\Omega_{\rm
in}} \! \diff^2 \theta' \, \Cov(\kappa; \vec{\theta}, \vec{\theta}') &
\cr
& \quad + \left( {A_{\rm in} \over A_{\rm out}} \right)^2
\int_{\Omega_{\rm out}} \!\! \diff^2 \theta \int_{\Omega_{\rm out}}
\!\! \diff^2 \theta' \, \Cov(\kappa; \vec{\theta}, \vec{\theta}') & \cr
& \simeq {c A \over 4 \rho} \; , & \autnum}
$$
where, we recall, $A$ is the area used.\footnote{*}{In principle, all
integrations should be performed in the whole plane. However, here we
suppose to perform integrations over a domain $\Omega$ of area $A$
large with respect to $\sigma_W^2$.} Obviously, the latter approximate
expression holds when $A_{\rm in} / A_{\rm out} \ll 1$. Curiously,
this result does not depend explicitly on the weight function $W$. The
derivation given in Appendix~C assumes that the weight function is of
the form of Eq.~(3), but a similar expression for the variance of $M$
is expected to hold in the more general case.

The results of this subsection can be clarified by a simple example.
Instead of introducing the weight function $W$, we consider the
unweighted Eqs.~(5) or (6) on small patches of the sky. For
simplicity, we refer to a square set $\Omega$ of length $L$ divided into
$s^2$ equal square patches: thus we expect ${\cal N} = \rho L^2 / s^2$
galaxies per patch. In this case the expected variance of $\gamma$ is
(see Paper I) $c / 4 {\cal N}$, and the variance of $\kappa$ is of the
same order of magnitude. The expected variance of $M$ is then $(c / 4
{\cal N}) \bigl( L^2 / s^2 \bigr)^2 s^2$, where the first factor is
the variance of $\kappa$ in every patch, the second factor is the area
of every patch (the square is necessary because we are dealing with
variances), and the third factor arises because we must add $s^2$
independent variables. The final result for the variance of $M$ is $c
L^2 / 4 \rho$, exactly as stated by Eq.~(32).

Here we may come back to the issue of the Poisson noise, only
mentioned at the beginning of Sect.~3. Strictly speaking, the relation
${\cal N} = \rho L^2 / s^2$ in the previous paragraph should be
replaced by $\langle {\cal N} \rangle = \rho L^2 / s^2$, with $\cal N$
following a Poisson distribution. We now consider the variance of
$\gamma$ as a function of $\cal N$. An estimate of the effect of the
Poisson noise can be obtained in the limit $\langle {\cal N} \rangle
\gg 1$ by expanding
$$
\Var(\gamma, {\cal N}) \simeq {c \over 4 \langle {\cal N} \rangle}
\left[1 - {{\cal N} - \langle {\cal N} \rangle \over \langle {\cal N}
\rangle} + \left( {{\cal N} - \langle {\cal N} \rangle \over \langle
{\cal N} \rangle} \right)^2 \right] \; .
\eqno \autnum
$$
Averaging over the ensemble thus yields
$$
\bigl\langle \Var(\gamma, {\cal N}) \bigr\rangle \simeq {c \over 4
\langle {\cal N} \rangle} \left[ 1 + {1 \over \langle {\cal N}
\rangle} \right] \; .
\eqno \autnum
$$
The effect of the Poisson noise is here contained in the second term
in brackets, which is negligibly small. Additional discussion is
postponed to the end of Appendix~B.

\titleb{The general case}
The situation is, in principle, quite similar to the weak lensing
limit, but, in practice, the calculations are much more difficult. If
the angular scale of $\kappa$ is much greater than the angular scale
of $W$, then we can prove that the mean value of the measured mass
distribution given by Eq.~(7) holds unchanged.

Difficulties in the calculation of the covariance of $\kappa$ mainly
arise from the form of the covariance of $g$ given by Eq.~(25),
because of the dependence on $\vec{\theta}$ and $\vec{\theta}'$ of the
first factor. However, if the lens has $\bigl| g_0(\vec{\theta})
\bigr| < 1$, the covariance given by Eq.~(25) is smaller than that of
the weak lensing limit of Eq.~(26) and thus we can consider all the
results given in the weak lensing regime as upper limits for the errors.

\titleb{Edge effects}
Finite boundaries introduce interesting effects, and make the errors
depend on the kernel $\vec{G}$ used in Eq.~(17). For simplicity we
take two different sets for $\vec{\theta}$ (see Fig.~1). The first set
is $\Omega'$, i.e.\ the observation area that includes all the lensed
galaxies used in the reconstruction.  The second set is $\Omega
\subset \Omega'$, i.e.\ the set where we measure $g(\vec{\theta})$. We
suppose that every point in $\Omega$ has a neighborhood with radius of
the order of the angular scale of $W$ completely enclosed in
$\Omega'$. This assumption greatly simplifies calculations and does
not have major practical consequences, except that it leads to
discarding a small strip $\Omega' \setminus \Omega$ around the
boundary $\partial \Omega'$ of $\Omega'$. With this hypothesis the
expected measured mass distribution is again given by Eq.~(7) as long
as $\vec{\theta} \in \Omega$. However this is strictly true only if we
choose correctly the mean mass distribution $\bar{\kappa}$ (see
Eq.~(17)).

\begfig 7.5 cm
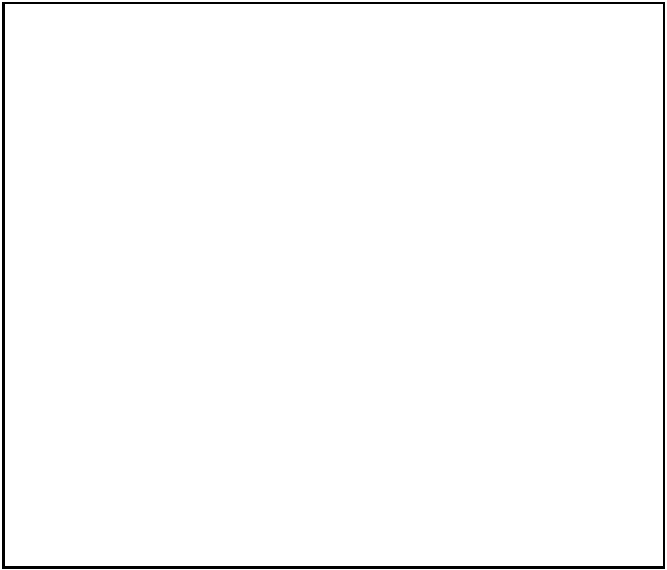
\figure{1}{ Sketch of the observation area used in the mass
reconstruction. }
\endfig

In general, the covariance depends on the kernel $\vec{G}$ used, and
in particular on its divergence-free component (see Appendix~C). [We
recall that a vector field $\vec{G}(\vec{\theta}, \vec{\theta}')$ can
be decomposed as $\vec{G} = \vec{G}' + \vec{G}''$, where $\vec{G}'$
has vanishing curl and $\vec{G}''$ has vanishing divergence (as usual,
in the above notation and decomposition, the emphasis is on the
variable $\vec{\theta}'$, since $\vec{\theta}$ is taken to be fixed).]
The result is
$$\eqalignno{
& \Cov(\kappa; \vec{\theta}, \vec{\phi}) = \Cov(\gamma; \vec{\theta},
\vec{\phi}) & \cr 
& \quad {} + \int_\Omega \! \diff^2 \theta' \!\! \int_\Omega \!
\diff^2 \, \phi' G_j''(\vec{\theta}, \vec{\theta}') \Cov(u;
\vec{\theta}', \vec{\phi}') G_j''(\vec{\phi}, \vec{\phi}') \; , &
\autnum \cr}
$$
where $\vec u$ is the quantity defined in Eq.~(11). The first term is
clearly independent of the kernel $\vec{G}$ used, while the second
term can be shown to be positive definite, i.e.
$$
\Cov(\kappa; \vec{\theta}, \vec{\theta}) \ge \Cov(\gamma; \vec{\theta},
\vec{\theta}) \; .
\eqno \autnum
$$
{\it Thus the error on $\kappa$ is minimized if a curl-free kernel
$\vec{G}$ is used}. This suggests that only curl-free kernels should
be used in weak lensing reconstructions. In fact, the kernels so far
judged to be ``good'' by means of simulations, all have vanishing curl
(see Sect.~5). For a curl-free kernel, such as the noise filtering
kernel given in Eqs.~(18--20), the result is independent of the kernel
used and of the set $\Omega$.

We now investigate the class of kernels $\vec{G}(\vec{\theta},
\vec{\theta}')$ that satisfy the following properties:
\medskip
\item{\it i.} $\vec{G}$ inverts Eq.~(11) when $\vec{u}(\vec{\theta})$
is measured with no error;
\item{\it ii.} $\vec{G}$ is curl-free.
\medskip
From the second property we can write
$$
\vec{G}(\vec{\theta}, \vec{\theta}') = -\nabla_{\vec{\theta}'} {\cal
G}(\vec{\theta}, \vec{\theta}') \; .
\eqno \autnum
$$
Thus, if $\vec{u}(\vec{\theta}') = \nabla \kappa(\vec{\theta}')$ we
find
$$\eqalignno{
\kappa(\vec{\theta}) - \bar{\kappa} &= -\int_\Omega \nabla {\cal
G}(\vec{\theta}, \vec{\theta}') \cdot \nabla \kappa(\vec{\theta}') \,
\diff^2 \theta' \cr
& = \int_\Omega \kappa(\vec{\theta}') \nabla^2 {\cal G}(\vec{\theta},
\vec{\theta}') \, \diff^2 \theta' \cr
& \phantom{{}=} - \int_{\partial\Omega}
\kappa(\vec{\theta}') \nabla {\cal G}(\vec{\theta}, \vec{\theta}')
\cdot \vec{n} \, \diff \theta' \; . & \autnum \cr}
$$
As in Sect.~2, $\vec{n}$ is the unit vector orthogonal to $\partial
\Omega$. The last relation shows that, in order to satisfy point
{\it i}., we must have
$$
\nabla^2 {\cal G}(\vec{\theta}, \vec{\theta}') = \delta(\vec{\theta}
- \vec{\theta}') - {1 \over A} \; . 
\eqno \autnum
$$
If $\kappa(\vec{\theta})$ is not known on the boundary
of $\Omega$, we should also consider:
$$
{\partial {\cal G}(\vec{\theta}, \vec{\theta}') \over \partial
\theta'_i} n_i = 0 \qquad \forall \vec{\theta}' \in \partial \Omega \;
.
\eqno \autnum
$$
In this case the kernel to be used is simply the one obtained from
${\cal G} = {\cal H}^{\rm SS}$, i.e.\ $\vec{G} = \vec{H}^{\rm SS}$
(cfr.\ Eqs.~(18--21)).  Otherwise, Eq.~(40) is to be dropped and the
kernel $\cal G$ is determined up to a term $\cal L$, where $\cal L$ is
a harmonic function ($\nabla^2 {\cal L} = 0$). This free function can
be used to dispose of the contribution that would arise from the
boundary term in Eq.~(38).

As, in general, the measured $\vec{u}$ field is not curl-free, the
inversion can only be approximate. The best inversion can thus be
found by searching for the function $\kappa(\vec{\theta})$ that
minimizes the functional
$$
{\cal S} = \int_\Omega \bigl\| \nabla \kappa(\vec{\theta}) -
\vec{u}(\vec{\theta}) \bigr\|^2 \, \diff^2 \theta \; .
\eqno \autnum
$$
If we vary the distribution $\kappa \mapsto \kappa + \delta\kappa$,
the functional would in general change to ${\cal S} + \delta{\cal S}$,
with
$$\eqalignno{
\delta{\cal S} &= 2 \int_\Omega \nabla (\delta \kappa) \cdot ( \nabla
\kappa - \vec{u} ) \, \diff^2 \theta & \cr
& = -2 \int_\Omega \delta \kappa ( \nabla^2 \kappa - \nabla \cdot
\vec{u} )\, \diff^2 \theta & \cr
& \phantom{{}=} + 2 \int_{\partial\Omega} \delta \kappa
(\nabla \kappa - \vec{u}) \cdot \vec{n} \, \diff \theta \; . &
\autnum \cr}
$$
By setting $\delta{\cal S} = 0$ we readily find the associated
Euler-Lagrange equation
$$
\nabla^2 \kappa(\vec{\theta}) = \nabla \cdot \vec{u}(\vec{\theta}) \;
.
\eqno \autnum
$$
This equation should be supplemented by
$$
\nabla \kappa(\vec{\theta}) \cdot \vec{n} = \vec{u}(\vec{\theta})
\cdot \vec{n} \qquad \forall \vec{\theta} \in \partial \Omega \; ,
\eqno \autnum
$$
unless we fix the value of $\kappa$ on $\partial\Omega$, so that the
boundary term in Eq.~(42) vanishes because $\delta\kappa = 0$ on
$\partial \Omega$. Equations~(43) and (44) define a Neumann boundary
problem equivalent to Eqs.~(39) and (40) above, in the sense that
$\cal G$ is precisely the Green function associated with it. This
clarifies the interesting properties of $\vec{H}^{\rm SS}$.

\titleb{Power spectrum}
In order to express in a simple manner the errors involved in the
reconstruction process, Seitz \& Schneider (1996) introduce a ``power
spectrum'' $P(k)$. In the weak lensing limit, their definitions are
$$\eqalignno{
& \Delta(\vec{k}) = {1 \over A} \int_\Omega \eul^{i \vec{k} \cdot
\vec{\theta}} \bigl( \kappa(\vec{\theta}) - \kappa_0(\vec{\theta})
\bigr) \, \diff^2 \theta \; , & \autnum \cr
& P(k) = \bigl\langle \Delta (\vec{k})
\Delta^*(\vec{k}) \bigr\rangle \; , & \autnum \cr}
$$
where the mean in Eq.~(46) is also over the various directions of
$\vec{k}$. As a result, the power spectrum $P(k)$ is simply the
variance of the complex map $\Delta(\vec{k})$, i.e.\ the Fourier
transform of the reconstruction error. Thus, for example, the value of
$P(0)$ is proportional to $\Var(M)$, the error on the total mass,
while its behavior for larger values of $k$ is related to the angular
scale of the weight function used.

Within our framework it is not difficult to evaluate the relevant
power spectrum. A simple calculation (see Appendix~D) for a Gaussian
weight function gives
$$
P(k) \simeq {c \over 4 \rho A} \exp \bigl[ - \sigma_W^2 k^2 \bigr] \; ,
\eqno \autnum
$$
i.e.\ a simple Gaussian with variance $1 / 2 \sigma_W^2$. One might
anticipate a significant contribution to the power spectrum coming
from the error associated with the difference between $\langle \kappa
\rangle$ and $\kappa_0$, but we argue in Appendix~D that such contribution
is negligible in the weak lensing limit.

\titlea{Comparison with numerical simulations}
In this section we compare our predictions with the results obtained
by Seitz \& Schneider (1996) from numerical simulations. Simulations
start by defining a lens mass distribution and a random sample of
source galaxies. Each galaxy is traced to the lens plane and the
reduced shear $g$ is then calculated from the observed ellipticities
using Eq.~(5). Finally the shear map is inverted into the lens mass
distribution $\kappa$ using various methods. For $\Omega$ Seitz \&
Schneider take a square $7.5^\prime \times 7.5^\prime$. Source
galaxies have random orientations and their ellipticities follow
truncated Gaussian distributions. Simulations have been performed with
three different variances for $\chi^{\rm s}$: $c_1 = 0.069109$, $c_2 =
0.13323$ and $c_3 = 0.19689$.

The reconstruction method used by  Seitz \& Schneider is similar
to the one described in Sect.~2, with the following differences:
\medskip
\item{\it i.} Their weight function is not invariant upon translations
because it is a Gaussian of argument $\| \vec{\theta} - \vec{\theta}'
\|$ with the variance depending on $\vec{\theta}$.
\item{\it ii.} An {\it outer smoothing\/} is added to the final lens
mass distribution.
\medskip
The first point is a device introduced in order to have better
resolution in the stronger parts of the lens. The second point is used
in order to have a smooth lens distribution from a discrete map of
$\kappa$.

Our result (47) for the power spectrum can be easily generalized in order
to take into account the outer smoothing:
$$
P(k) \simeq {c \over 4 \rho A} \exp \bigl[ - \bigl(\sigma_W^2 +
\sigma_s^2 \bigr) k^2 \bigr] \; .
\eqno \autnum
$$
Here $\sigma_s^2$ is the variance associated with the outer smoothing.
Note however that the expression given above does {\it not\/} take
into account the variable-scale smoothing used by Seitz \& Schneider.
Even if this result has been derived with some approximations (weak
lensing limit, large area $A$ of $\Omega$, fixed inner smoothing
$\sigma_W$), a comparison with the simulations shows that Eq.~(48) can
reproduce the main features of the simulated power spectrum.

\begfig 16cm
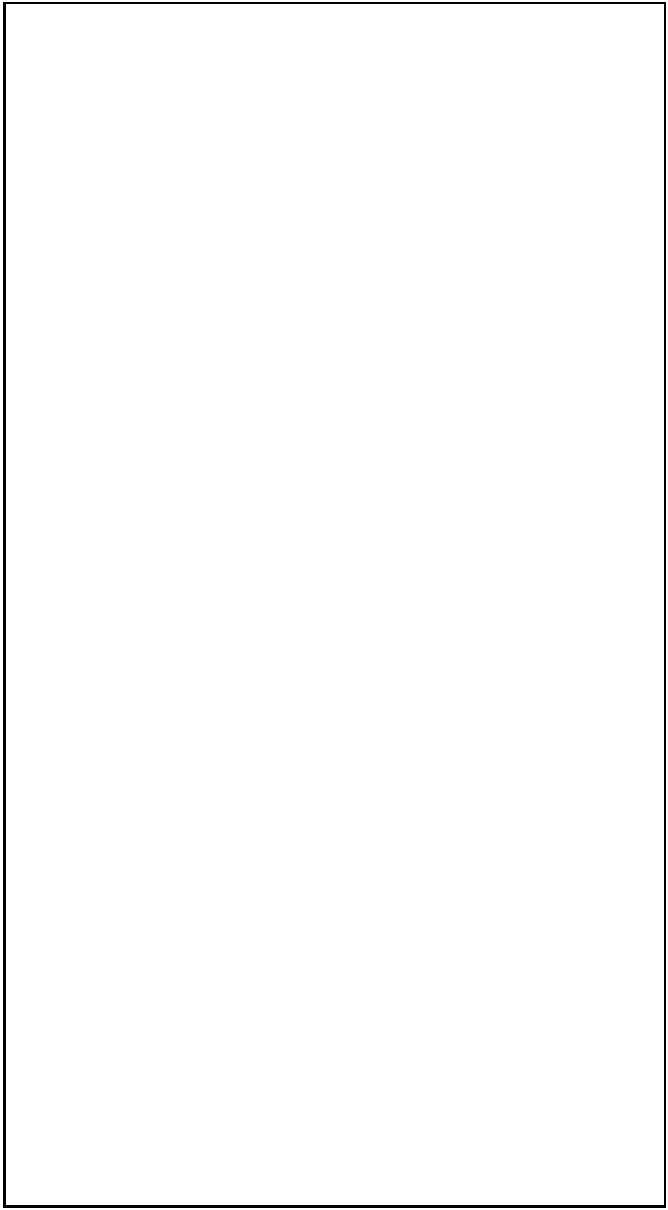
\figure{2}{Comparison of predicted power spectrum (dashed lines) with
measured power spectrum (solid lines) for the simulations made by
Seitz \& Schneider (1996). All frames refer to a source population
characterized by $c = 0.069109$. Top frame: $\rho A = 80$, $\sigma_W =
0.212^\prime$, $\sigma_s = 0.0778^\prime$. Middle frame: $\rho A =
50$, $\sigma_W = 0.177^\prime$, $\sigma_s = 0^\prime$. Bottom frame:
$\rho A = 50$, $\sigma_W = 0.240^\prime$, $\sigma_s =
0.0778^\prime$. }
\endfig

Figure~2 shows the results of simulations together with the
power spectrum predicted by Eq.~(48). Thus Eq.~(48) underestimates the
error. This difference can be attributed to the following factors:
\medskip
\item{1.} The weight function $W$ considered is not precisely of the
form of Eq.~(3), because of the change of normalization near $\partial
\Omega$. This last factor should increase the variance of $\kappa$
near the boundary of $\Omega$ (the variance of $\gamma$, with direct
influence on $\kappa$, should double near a side of $\Omega$ and
quadruple near a corner; cf.\ top-right frame of Fig.~10 in Seitz \&
Schneider 1996).
\item{2.} The weight function $W$ is not of the form of Eq.~(3) also
because of differential smoothing.
\item{3.} The constant term $\bar{\kappa}$ has not been estimated
properly (see first paragraph of Sect.~4.3). In principle, this should
be traced to a counterpart in $P(0)$, but for finite sets $\Omega$
there is an additional term in $P(k)$.
\item{4.} The set $\Omega$ is not the whole plane.
\item{5.} The lens is not weak (see Eq.~(25) and the extra
contribution in Eq.~(D4)).
\item{6.} Poisson noise is associated with the finite number of
source galaxies.
\item{7.} The population of source galaxies is characterized by
sizable $c$.
\medskip
In spite of these limitations, the general behavior of $P(k)$ is
reasonably well reproduced by Eq.~(48). In particular, the maximum of
the simulated points corresponds exactly to the maximum of our
theoretical curve.

\titleb{Curl-free kernels}
In order to check the result of Eq.~(35), we have considered different
kernels used by various authors and we have compared our predictions
with other aspects of the numerical simulations performed by Seitz \&
Schneider (1996).

The first kernel considered is the noise-filtered ``SS-inversion''
(Seitz \& Schneider 1996) described above in Eqs.~(18--20). This
method has been especially designed to reduce the statistical errors
and performs very well in simulations. In fact, as stated in Eq.~(21),
this kernel is curl-free.

Another kernel considered is the ``S-inversion'' (see Schneider
1995). Simulations show that errors of the S-inversion are nearly the
same as for the SS-inversion. The S-inversion operates by averaging
over radial paths made of two segments. Inside the segments the kernel
is easily shown to be curl-free.

The last kernel is the so called ``B-inversion'' (Bartelmann 1995; see
also Squires \& Kaiser 1996). From the results of the simulations it
is clear that the B-inversion leads to larger errors on the map
distribution. This behavior is once again explained by Eq.~(35), since
the B-inversion kernel is {\it not\/} curl-free. Notice that in this
case it is difficult to estimate analytically the exact error on
$\kappa$.

\acknow{We would like to thank Peter Schneider for several comments
and suggestions that have helped us improve the paper. This work has
been partially supported by MURST and by ASI of Italy.}

\appendix{A. Notation}

We collect here the main symbols used in this paper.
\medskip
\begingroup
\itemindent=35pt
\item{$X^{\rm s}$} Superscript s identifies {\it source\/} (unlensed)
quantities. 
\item{$X_i$} Subscripts refer to the complex representation, e.g.\ $\chi =
\chi_1 + \imag \chi_2$ or, when indicated, to the vector
representation, e.g.\ $\vec{\theta} = (\theta_1, \theta_2)$.
\item{$\vec{\theta}^{\rm s}$, $\vec{\theta}$} Unlensed and observed
position of a point source.
\item{$\Sigma_{\rm c}$} Critical density.
\item{$\Sigma(\vec{\theta})$} Projected mass distribution of the lens.
\item{$\kappa(\vec{\theta})$} Dimensionless mass distribution:
$\kappa(\vec{\theta}) = \Sigma(\vec{\theta} / \Sigma_{\rm c}$.
\item{$Q^{\rm s}$, $Q$} Unlensed and observed quadrupole
moment of an extended image.
\item{$\chi^{\rm s}$, $\chi$} Unlensed and observed (complex)
ellipticity of an extended image: $\chi^{\rm s} = \bigl(Q^{\rm s}_{11}
- Q^{\rm s}_{22} + 2 \imag Q^{\rm s}_{12} \bigr) / \bigl( Q^{\rm
s}_{11} + Q^{\rm s}_{22} \bigr)$, and similarly for $\chi$.
\item{$\gamma$} Complex shear.
\item{$g$} Complex reduced shear: $g = \gamma / (1 - \kappa)$.
\item{$c$} Covariance of the source ellipticity for an isotropic
distribution: $\bigl\langle \chi^{\rm s}_i \chi^{\rm s}_j
\bigr\rangle = c \delta_{ij}$. 
\item{Id, $\delta_{ij}$} Identity matrix.
\endgroup

\appendix{B. The shear}
In this Appendix we derive Eq.~(23) and Eq.~(25) of Sect.~3. We assume
that the reduced shear $g$ has been measured through
Eq.~(5). Calculations based on Eq.~(6) are very similar. As explained
in Appendix~A of Paper~I, the mean value of $g$ obeys the relation
$$
\sum_{n=1}^N W\bigl( \vec{\theta}, \vec{\theta}^{(n)} \bigr) \chi^{\rm
s}  \left( \bigl\langle \chi^{(n)} \bigr\rangle , \bigl\langle 
g(\vec{\theta}) \bigr\rangle \right) = 0 \; .
\eqno \xautnum{B}
$$
Here $\bigl\langle \chi^{(n)} \bigr\rangle$ is the mean value of
$\chi$ in $\vec{\theta}^{(n)}$ and thus depends on
$g_0\bigl(\vec{\theta}^{(n)} \bigr)$. We now assume that
$g_0(\vec{\theta})$ does not change significantly on the angular scale
of $W(\vec{\theta}, \vec{\theta}')$. This implies that we can expand
the previous equation to first order in $g_0$. We choose, as starting
point, the value $g_0 = g_*$ given
by
$$
g_*(\vec{\theta}) = \sum_{n=1}^N W \bigl( \vec{\theta},
\vec{\theta}^{(n)} \bigr) g_0 \bigl( \vec{\theta}^{(n)} \bigr) \; .
\eqno \xautnum{B}
$$
Then we find easily
$$\eqalignno{
&\sum_{n=1}^N W \chi^{\rm s}  \left( \chi_*, \bigl\langle
g(\vec{\theta}) \bigr\rangle \right) & \cr 
& \quad {} + \sum_{n=1}^N W \left. {\partial \chi^{\rm s} \over
\partial \chi} \right|_{\chi = \chi_*} {\partial \chi_* \over \partial
g_*} \left( g_0 \bigl(\vec{\theta}^{(n)} \bigr) - g_* \right) = 0 \; ,
& \xautnum{B}}
$$
where $\chi_*$ is the expected value of the ellipticity when the
reduced shear is equal to $g_*$. Equation (B3) has the obvious
solution $\bigl\langle g(\vec{\theta}) \bigr\rangle =
g_*(\vec{\theta})$, because the first term, $\chi^{\rm s} (\chi_*,
g_*)$, vanishes by definition and the second vanishes due to the
choice of $g_*$ (notice that the partial derivatives in the latter do
not depend on $\vec{\theta}^{(n)}$). When moving to a continuous
description, we have to calculate the average expected value of $g$
over all possible positions $\bigl\{ \vec{\theta}^{(n)} \bigr\}$ of
the source galaxies
$$\eqalignno{
\bigl\langle g(\vec{\theta}) \bigr\rangle = {} & {1 \over A^N}
\int_\Omega \diff^2 \theta^{(1)} \int_\Omega \diff^2 \theta^{(2)}
\cdots \int_\Omega \diff^2 \theta^{(N)} \times {} & \cr
& {} \times \sum_{n=1}^N W \bigl( \vec{\theta}, \vec{\theta}^{(n)}
\bigr) g_0 \bigl( \vec{\theta}^{(n)} \bigr) \; . & \xautnum{B} \cr}
$$
The weight function $W \bigl( \vec{\theta}, \vec{\theta}^{(n)} \bigr)$
depends on all $\bigl\{\, \vec{\theta}^{(1)}, \vec{\theta}^{(2)},
\dots, \vec{\theta}^{(N)} \,\bigr\}$ because of the adopted
normalization. However, in the limit $N \gg 1$, the weight function $W
\bigl( \vec{\theta}, \vec{\theta}^{(n)} \bigr)$ can be considered to
depend only on $\vec{\theta}^{(n)}$ alone, so that the above integral
can be approximated by
$$\eqalignno{
\bigl\langle g(\vec{\theta}) \bigr\rangle & \simeq {1 \over A}
\sum_{n=1}^N \int_\Omega \diff^2 \theta^{(n)} \, W \bigl(
\vec{\theta}, \vec{\theta}^{(n)} \bigr) g_0 \bigl( \vec{\theta}^{(n)}
\bigr) & \cr 
& = \rho \int_\Omega \diff^2 \theta' \, W(\vec{\theta}, \vec{\theta}')
g_0(\vec{\theta}') \; . & \xautnum{B} \cr}
$$
This proves Eq.~(23). Our result simply states that the use of the
first order expansion in $g$ reduces every mean to a weighted
arithmetic mean.

Calculations for the covariance of $g$ are much more difficult but
basically repeat those given for the unweighted situation in App.~A.1
of Paper~I. In particular, if we call $F \bigl( \bigl\{
\vec{\theta}^{(n)}, \chi^{(n)} \bigr\}, g(\vec{\theta}) \bigr)$ the
function defined in the l.h.s.\ of Eq.~(5), we have 
$$\eqalignno{
& \Cov (g; \vec{\theta}, \vec{\theta}') = B^{-1}(\vec{\theta}) &
\cr
& \quad \biggl[ \sum_{n=1}^N A^{(n)}(\vec{\theta}) \Cov\bigl(
\chi^{(n)} \bigr) A^{(n)T} (\vec{\theta}') \biggr] \bigl(
B^{-1}(\vec{\theta}') \bigr)^T \; , & \xautnum{B} \cr} 
$$
where
$$\eqalignno{
& B(\vec{\theta}) = {\partial F \over \partial g(\vec{\theta})} \; , &
\xautnum{B} \cr
& A^{(n)}(\vec{\theta}) =  {\partial F \over \partial \chi^{(n)}} \; . &
\xautnum{B} \cr}
$$
All functions have to be calculated in the mean value of their
arguments. Some calculations then lead to
$$
B(\vec{\theta}) \simeq {2 \over 1 - \bigl| \langle g(\vec{\theta})
\rangle \bigr|^2} \, {\rm Id} \; .
\eqno \xautnum{B}
$$
The term in brackets in Eq.~(B6) can be written in the form
$$\eqalignno{
& \sum_{n=1}^N A^{(n)}(\vec{\theta}) \Cov\bigl( \chi^{(n)} \bigr)
A^{(n)T} (\vec{\theta}') & \cr
& \quad = c \sum_{n=1}^N W\bigl( \vec{\theta}, \vec{\theta}^{(n)}
\bigr) W\bigl( \vec{\theta}', \vec{\theta}^{(n)} \bigr) + \hbox{\rm
linear terms} \; . & \xautnum{B} \cr}
$$
Here ``linear terms'' means additional terms linear with respect to
the quantity $\bigl[ g_0\bigl( \vec{\theta}^{(n)} \bigr) - g_*
\bigr]$, based on the same expansion defined by~(B2).

By averaging over the source positions and by moving to a continuous
description (basically following the same steps indicated in (B4) and
(B5)), we thus obtain Eq.~(25). Notice that the ``linear terms'' in
Eq.~(B10) do not give any contribution when averaged over the source
positions. We stress that the results stated here are valid only if
the weight function is even (property 1 of Sect.~2.1).

Finally, we point out that the approximation that takes us from (B4)
to (B5) is precisely associated with neglecting the Poisson noise.

\appendix{C. The lens mass distribution}
In this Appendix we will derive Eq.~(7) and the results stated in
Sect.~4, assuming a weight function invariant upon translations (case
2 of Sect.~2.1).

\xtitleb{C}{Weak lensing limit}
Calculations in the weak lensing limit are not difficult. As explained
in Sect.~2.2, we can use either Eq.~(8) or Eqs.~(11) and (12) to
convert the reduced shear into the mass distribution.

In the case of Eq.~(8) we can write
$$
\langle \kappa \rangle = D_i \star \langle \gamma_i \rangle = D_i
\star ( \rho W \star \gamma_{0i} ) \; .
\eqno \xautnum{C}
$$
Here the star denotes convolution, while $\gamma_{0i}$ is the
component $i$ of the true shear map $\gamma_0$. The second step is
justified by Eq.~(23) applied in the weak lensing limit. By using the
associative and commutative properties of the convolution and by noting
that $\kappa_0 = D_i \star \gamma_{0i}$, we can write
$$\eqalignno{
\langle \kappa \rangle &= D_i \star \rho W \star \gamma_{0i} & \cr
& = \rho W \star D_i \star \gamma_{0i} = \rho W \star \kappa_0 \; . &
\xautnum{C} \cr}
$$
This proves Eq.~(7).

About the covariance of $\kappa$ we can write
$$\eqalignno{
& \Cov(\kappa; \vec{\theta}, \vec{\phi}) = \int \diff^2 \theta'
\int \diff^2 \phi' \, D_i(\vec{\theta} - \vec{\theta}') \times {} & \cr
& \quad {} \times \Cov(\gamma; \vec{\theta}' - \vec{\phi}')
D_i(\vec{\phi} - \vec{\phi}') \; . & \xautnum{C} \cr}
$$
As $D_i$ and $\Cov(\gamma)$ are even functions, this is simply a
double convolution, and thus the result depends only on the difference
between $\vec{\theta}$ and $\vec{\phi}$. Therefore we can write for
$\Cov(\kappa; \vec{\theta}, \vec{\phi}) = \Cov(\kappa; \vec{\theta} -
\vec{\phi})$ the expression
$$
\Cov(\kappa) = D_i \star \Cov(\gamma) \star D_i = \Cov(\gamma) \;
,
\eqno \xautnum{C}
$$
i.e.\ Eq.~(31). Here we have used again the commutative property of
convolutions and the relation $D_i \star D_i = \delta$ given by
Eqs.~(8) and (10). [Hereafter $\delta$ means the Dirac delta
distribution.]

The results are the same if we use Eqs.~(11) and (12). In fact we
can write Eq.~(11) as a convolution between $\gamma_i$ and the operator
$$
T_{ij}(\vec{\theta}) = \left( \matrix{\delta_{,1}(\vec{\theta}) &
\delta_{,2}(\vec{\theta}) \cr -\delta_{,2}(\vec{\theta}) &
\delta_{,1}(\vec{\theta})} \right) \; ,
\eqno \xautnum{C}
$$
where $\delta_{,i}(\vec{\theta}) = \partial \delta(\vec{\theta}) /
\partial \theta_i$. Thus we are allowed to use the properties of
convolutions. It is obvious then that the convolution with the weight
function $W$ in $\langle \gamma \rangle$ can be moved to the true lens
distribution $\kappa_0$, and we find again the result of Eq.~(C2). The
covariance matrix of $\vec{u}$ can be calculated using the operator
(C5). As a result, we find
$$
\Cov(u) = - \nabla^2 \Cov(\gamma) \; .
\eqno \xautnum{C}
$$
We then have 
$$\eqalignno{
\Cov(\kappa) & = - H^{\rm KS}_i \star \nabla^2 \Cov(\gamma) \star
H^{\rm KS}_i & \cr 
& = - \nabla^2 (H^{\rm KS}_i \star H^{\rm KS}_i) \star \Cov(\gamma) \;
. & \xautnum{C} \cr}
$$
A simple calculation shows that $-\nabla^2 (H^{\rm KS}_i \star
H^{\rm KS}_i)(\vec{\theta}) = \nabla^2 \bigl( \ln \| \vec{\theta} \| /
2 \pi \bigl) = \delta(\vec{\theta})$. This leads again to Eq.~(31).

Now let us prove Eq.~(32). From Eq.~(C4) and Eq.~(26) we find
$$
\Cov(\kappa) = {c \rho \over 4} W \star W \; .
\eqno \xautnum{C}
$$
In the limit $A_{\rm in}/A_{\rm out} \ll 1$ explained in Sect.~4.1,
the main contribution to the variance of $M$ derives from a double
integration of $\Cov(\kappa)$. A simple change of variables gives
$$\eqalignno{
\Var(M) & = {c \rho \over 4} \int \diff^2 \theta \int \diff^2 \theta' \,
(W \star W)(\vec{\theta}') & \cr
& = {c \rho A \over 4} \hat{W}(\vec{0}) \hat{W}(\vec{0}) = {c A \over
4 \rho} \; . & \xautnum{C} \cr}
$$
Here $\hat{W}$ is the Fourier transform of $W$ and the last equality
holds because of the normalization (22) of the weight function.

\xtitleb{C}{The general case}
In the general case we restrict ourselves to estimating the mean value
of the lens distribution because calculations for the covariance are
too difficult. Under the hypothesis that the angular scale of $W$ is
much smaller than the angular scale of $\kappa$ (or $g$), the
situation is much like that of the weak lensing limit. As shown in
Appendix~B, this basically implies that all averages are weighted
arithmetic averages. Simple calculations show that we have $\langle
\tilde{\vec{u}} \rangle = \tilde{\vec{u}}_0 \star \rho W$, and hence
$\langle \tilde{\kappa} \rangle = \tilde{\kappa}_0 \star
\rho W$. As usual the assumed ordering of scale lengths leads again to
Eq.~(C2).

\xtitleb{C}{Edge effects}
For simplicity we refer to $\bar{\kappa} = 0$. We rewrite Eqs~(17),
(11) and (10) with a different notation
$$\eqalignno{
& \pi_\Omega \kappa = G_i u_i \; , & \xautnum{C} \cr
& u_i = T_{ij} \star \gamma_j \; , & \xautnum{C} \cr
& \gamma_i = D_i \star \kappa \; . & \xautnum{C}}
$$
Here $\pi_\Omega$ is the characteristic operator for the set $\Omega$:
$$
(\pi_\Omega f)(\vec{\theta}) = \cases{ f(\vec{\theta}) & if
$\vec{\theta} \in \Omega \; ,$ \cr 0 & otherwise $\; .$ }
\eqno \xautnum{C} 
$$
Equation (C10) is equivalent to Eq.~(17) with $\bar{\kappa} = 0$ if we
redefine the kernel $\vec{G}(\vec{\theta}, \vec{\theta}')$ for every
$\vec{\theta}$ and $\vec{\theta}'$ so that $\vec{G}(\vec{\theta},
\vec{\theta}') = 0$ if either $\vec{\theta} \notin \Omega$ or
$\vec{\theta}' \notin \Omega$. With this simple definition we can
extend the integration domain (usually $\Omega$) to the whole plane.
Notice that while $T_{ij}$ and $D_i$ are used in convolutions, $G_i$
is a generic linear operator. From these equations we have
$$
\pi_\Omega \kappa = G_i u_i = G_i (T_{ij} \star D_j \star \kappa)
\; ,
\eqno \xautnum{C}
$$
and thus we find the identity
$$
G_i T_{ij} \star D_j = \pi_\Omega \; .
\eqno \xautnum{C}
$$
Equation (12) with the new notation is
$$
\kappa = H^{\rm KS}_i \star u_i \; .
\eqno \xautnum{C}
$$
This, together with Eqs.~(C11) and (C12), gives us another identity:
$$
u_i = T_{ij} \star D_j \star H^{\rm KS}_k \star u_k \; .
\eqno \xautnum{C}
$$
Using Eqs.~(C10), (C12), and the relation $\langle u_i \rangle = \rho
W \star u_{0i}$, we can easily obtain the mean value for measures of
$\kappa$:
$$\eqalignno{
\langle \kappa \rangle & = G_i \langle u_i \rangle = G_i \rho W \star
u_{0i} & \cr
& = G_i T_{ij} \star D_j \star H^{\rm KS}_k \star u_{0k} \star \rho W
\cr
& = \pi_\Omega H^{\rm KS}_k \star u_{0k} \star \rho W = \pi_\Omega
\rho W \star \kappa_0 \; . & \xautnum{C}}
$$
As usual, subscript $0$ indicates the true value of a quantity. This
equation, rewritten in the more standard notation, is Eq.~(7) for
$\vec{\theta} \in \Omega$.

Let us calculate the covariance of $\kappa$. First of all note that,
while Eq.~(C14) implies Eq.~(C15), from Eq.~(C17) we cannot deduce that
$T_{ij} \star D_j \star H^{\rm KS}_k$ is the identity. This happens
because the two components of $u_i$ are not functionally independent,
as one can see from the relation $\nabla \wedge \vec{u} = 0$. In fact,
using Fourier transforms it is easy to prove that the operator $R_{ik}
= T_{ij} \star D_j \star H^{\rm KS}_k$ selects the curl-free component
of a vector field. Its Fourier transform is
$$
\hat{R}_{ij}(\vec{k}) = {k_i k_j \over \| \vec{k} \|^2} \; .
\eqno \xautnum{C}
$$

From (C10) we have
$$
\Cov(\kappa) = G_j \Cov(u) G_j \; .
\eqno \xautnum{C}
$$
Every vector field can be written as the sum of two vector fields, of
which one is curl-free and the other is divergence-free. Hence, if we
consider $\vec{G}(\vec{\theta}, \vec{\theta}')$ a vector field with
respect to $\vec{\theta}'$, we can write
$$
\vec{G}(\vec{\theta}, \vec{\theta}') = \vec{G}'(\vec{\theta},
\vec{\theta}') + \vec{G}''(\vec{\theta}, \vec{\theta}') \; ,
\eqno \xautnum{C}
$$
where
$$\eqalignno{
& \nabla_{\vec{\theta}'} \wedge \vec{G}'(\vec{\theta}, \vec{\theta}')
= 0 \; , & \xautnum{C} \cr
& \nabla_{\vec{\theta}'} \cdot \vec{G}''(\vec{\theta}, \vec{\theta}')
= 0 \; . & \xautnum{C} \cr}
$$
Thus $\vec{G}'$ and $\vec{G}''$ can be written as the gradient and the
``curl'' of two scalar fields:
$$\eqalignno{
& \vec{G}'(\vec{\theta}, \vec{\theta}') = \nabla_{\vec{\theta}'}
s'(\vec{\theta}, \vec{\theta}') \; , & \xautnum{C} \cr
& \vec{G}''(\vec{\theta}, \vec{\theta}') = \nabla_{\vec{\theta}'}
\wedge s''(\vec{\theta}, \vec{\theta}') = \left( \matrix{
s''_{,2}(\vec{\theta}, \vec{\theta}') \cr -s''_{,1}(\vec{\theta},
\vec{\theta}') } \right) \; . & \xautnum{C} \cr} 
$$
There is some freedom in the choice of $\vec{G}'$ and $\vec{G}''$ (or
equivalently of $s'$ and $s''$). However, it is always possible to
choose $\vec{G}'$ and $\vec{G''}$ so that they vanish for $\|
\vec{\theta}' \| \rightarrow \infty$. With the decomposition (C21) we
have $G_i R_{ij} = G'_j$ and thus 
$$
\Cov(\kappa) = (G_i R_{ij} + G''_j) \Cov(u) (G_k R_{kj} + G''_j)
\eqno \xautnum{C}
$$
Recalling now the definition of $R_{ij}$ and using Eq.~(C15) we find
$$
\Cov(\kappa) = (\pi_\Omega H^{\rm KS}_j + G''_j) \Cov(u) (\pi_\Omega
H^{\rm KS}_j + G''_j)
\eqno \xautnum{C}
$$
If the kernel $\vec{G}$ has vanishing curl, then $\vec{G}'' = 0$ and
we find the final result
$$\eqalignno{
& \Cov(\kappa; \vec{\theta}, \vec{\phi}) = \int \diff^2 \theta' \int
\diff^2 \phi' \,  H^{\rm KS}_j(\vec{\theta} - \vec{\theta}') \times {}
& \cr & \quad {} \times \Cov(u; \vec{\theta}' - \vec{\phi}')
H^{\rm KS}_j(\vec{\phi} - \vec{\phi}') \qquad \hbox{\rm for }
\vec{\theta}, \vec{\phi} \in \Omega \; . & \xautnum{C} \cr}
$$
In general however we must evaluate three additional terms. Two of
them are of the form
$$\eqalignno{
& G''_j \Cov(u) G'_j = \int \diff^2 \theta' \int \diff^2 \phi'\,
\vec{G}''(\vec{\theta}, \vec{\theta}') \times {} & \cr
& \quad {} \times \Cov(u; \vec{\theta}' - \vec{\phi}')
\cdot \nabla_{\vec{\phi}'} s'(\vec{\phi}, \vec{\phi}') \; , &
\xautnum{C} \cr}
$$
where we have used Eq.~(C24). By the change of variable $\vec{\phi}'
\mapsto \vec{\theta}' - \vec{\phi}'$ and after integrating by parts we
find
$$\eqalignno{
& G''_j \Cov(u) G'_j = \int \diff^2 \theta' \int \diff^2 \phi'\,
\nabla_{\vec{\theta}'} \cdot \vec{G}''(\vec{\theta}, \vec{\theta}')
\times {} & \cr 
& \quad {} \times \Cov(u; \vec{\theta}' - \vec{\phi}')
s''(\vec{\phi} - \vec{\phi}') = 0 \; , & \xautnum{C} \cr}
$$
where the last relation holds in virtue of Eq.~(C23). We finally
rewrite Eq.~(C27) in a simplified form:
$$
\Cov(\kappa) = \pi_\Omega H^{\rm KS}_j \Cov(u) \pi_\Omega H^{\rm KS}_j
+ G''_j \Cov(u) G''_j
\eqno \xautnum{C}
$$
Here the first term is independent of the specific kernel $\vec{G}$
used, while the second term depends only on $\vec{G}''$. As, by
definition, $\Cov(u)$ is positive definite, the last term in Eq.~(C31)
is also positive definite. In other words if $\vec{G}''
\neq 0$ the error on $\kappa$ will increase.

\appendix{D. Power spectrum}
The power spectrum reported in Eq.~(47) can be deduced from the
expression of the mean and covariance of the measured mass
distribution. In particular we have
$$\eqalignno{
& \bigl\langle \Delta_i(\vec{k}) \bigr\rangle = {1 \over A} \int_\Omega
\Bigl( \eul^{\imag \vec{k} \cdot \vec{\theta}} \Bigr)_i \bigl(
\bigl\langle \kappa(\vec{\theta}) \bigr\rangle -
\kappa_0(\vec{\theta}) \bigr) \, \diff^2 \theta \; , & \xautnum{D} \cr
& \Cov_{ij}(\Delta; \vec{k}, \vec{k}') = {1 \over A^2} \int_\Omega \diff^2
\theta \int_\Omega \diff^2 \theta' \, \Bigl( \eul^{\imag \vec{k} \cdot
\vec{\theta}} \Bigr)_i \times {} & \cr
& \quad {} \times \Bigl( \eul^{\imag \vec{k}' \cdot \vec{\theta}'}
\Bigr)_j \Cov(\kappa; \vec{\theta}, \vec{\theta}') \; . & \xautnum {D}
\cr}
$$
The subscripts in the exponentials denote real ($i, j = 1$) and
imaginary ($i,j = 2$) parts. The power spectrum $P(k)$ is directly
related to the covariance of $\Delta$. In fact, we have
$$
P(k) = \Cov_{ii}(\Delta; \vec{k}, \vec{k}) + \bigl\langle
\Delta_i(\vec{k}) \bigr\rangle \bigl\langle \Delta_i(\vec{k})
\bigr\rangle \; , 
\eqno \xautnum{D}
$$
with summation implied on $i$ and mean over all directions of
$\vec{k}$.\ For a large set $\Omega$ we can perform integrations
over the whole plane. Thus we find
$$
P(k) = {1 \over A} \widehat{\Cov}(\kappa; \vec{k}) + {1 \over A^2}
\bigl| \hat{\kappa}_0(\vec{k}) \bigr|^2 \bigl| \rho \hat{W}(\vec{k}) -
1 \bigr|^2 \; .
\eqno \xautnum{D}
$$
Here, as usual, hats indicate Fourier transform. In the weak lensing
limit the second term of this expression can be dropped. Hence, if $W$
is a Gaussian of the form of Eq.~(3) we find Eq.~(47). [Note that no
averaging over the direction of $\vec{k}$ is needed in the weak
lensing limit if the weight function has the form (3).]

\begref{References}
%
\ref Bartelmann M., 1995, A\&A 303, 643
%
\ref Bartelmann M., Narayan R., Seitz S., Schneider P., 1996, ApJ 464,
L115
%
\ref Fahlman G.G., Kaiser N., Squires G., Woods D., 1994, ApJ 437, 56
%
\ref Falco E.E., Gorenstein M.V., Shapiro I.I., 1985, ApJ 289, L1
\ref Kaiser N., 1995, ApJ 493, L1
%
\ref Kaiser N. \& Squires J., 1993, ApJ 404, 441
%
\ref Kneib J.P, Ellis R.S., Smail I., Couch W.J., Sharples R.M., 1996,
ApJ 417, 643
%
\ref Lombardi M., Bertin G., 1998, A\&A 330, 791 (Paper~I)
%
\ref Lucy L.B, 1994, A\&A 289, 983
%
\ref Schneider P., 1996, A\&A 302, 639
%
\ref Schneider P., Ehlers J., Falco E.E., 1992, Gravitational Lenses,
Springer, Heidelberg
%
\ref Schneider P., Seitz C., 1995, A\&A 294, 411
%
\ref Seitz C., Schneider P., 1995, A\&A 297, 287
%
\ref Seitz S., Schneider P., 1996, A\&A 305, 383 
%
\ref Smail I., Ellis R.S., Fitchett M.J., Edge A.C., 1994, MNRAS 237,
257
\ref Squires G., Kaiser N., 1996, ApJ 473, 65
\ref Tyson J.A., Valdes F., Wenk R.A., 1990, ApJ 349, L1
%
\ref Webster R.L., 1985, MNRAS 213, 871

\endref

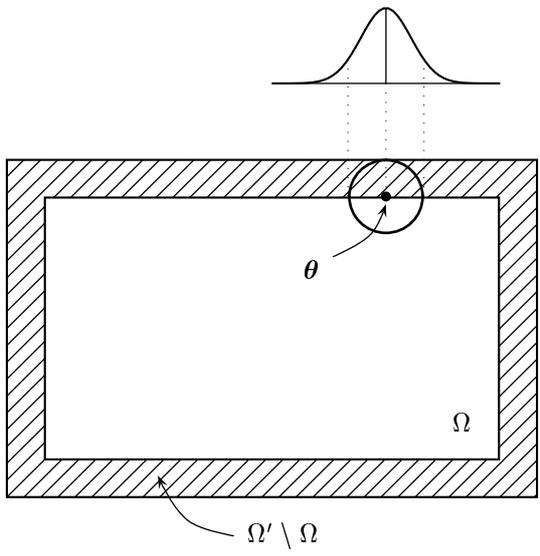
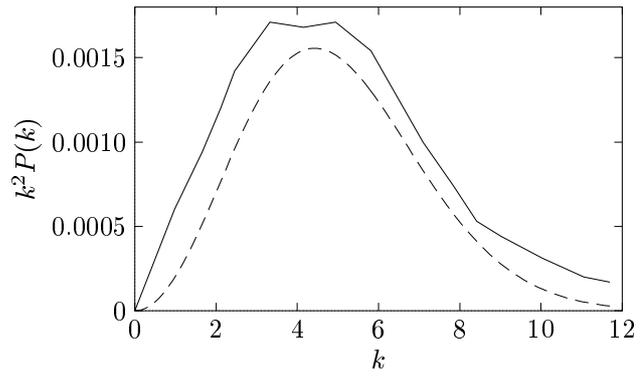
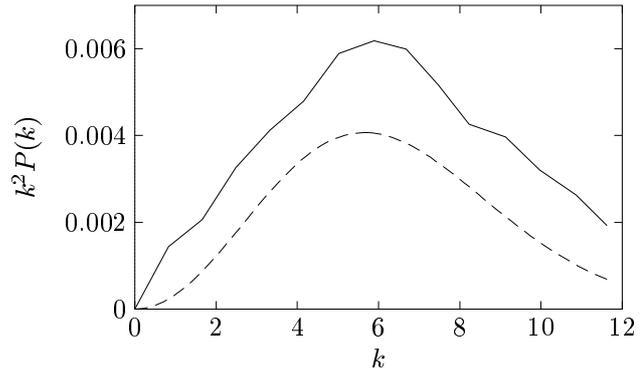
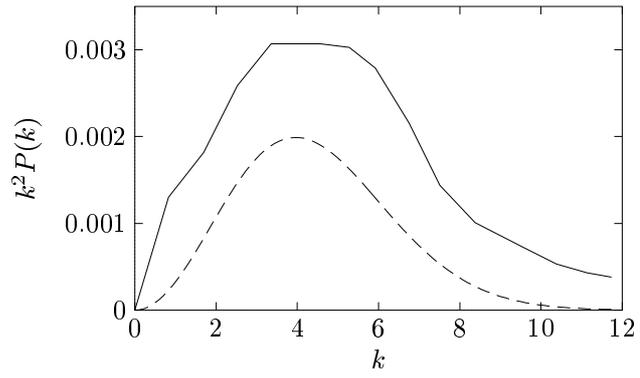
\bye